\documentstyle[12pt]{article}

\renewcommand{\baselinestretch}{1.5}

\def\beq{\begin{equation}}
\def\eeq{\end{equation}}
\def\bea{\begin{eqnarray}}
\def\nn{\nonumber \\ }
\def\eea{\end{eqnarray}}
\def\ds{\displaystyle}

\def\ms{\medskip}

\def\ni{\noindent}

\def\req#1{(\ref{#1})}
\def\ie{i.e. }
\def\eg{e.g. }

\def\rep{{\rm Re}\ }
\def\imp{{\rm Im}\ }

\begin{document}

%\hfill UB-ECM-PF 92/33

%\hfill December 1992

\vspace*{3mm}

\begin{center}

{\LARGE \bf
Dynamical Determination of the Metric Signature in Spacetime
of Nontrivial Topology}

\vskip4ex

\renewcommand\baselinestretch{0.8}
{\sc E. Elizalde}
\footnote{E-mail: eli @ ebubecm1.bitnet} \\
Department E.C.M. and I.F.A.E., Faculty of Physics,
University of  Barcelona, \\ Diagonal 647, 08028 Barcelona, \\
and Center for Advanced Studies, C.S.I.C., Cam\'{\i} de Santa B\`arbara, \\
17300 Blanes, Catalonia, Spain \\
{\sc S.D. Odintsov}\footnote{
On leave from Tomsk Pedagogical Institute, 634041 Tomsk, Russian
Federation. E-mail: odintsov @
ebubecm1.bitnet} and {\sc A. Romeo} \\
Department E.C.M., Faculty of Physics,
University of  Barcelona, \\  Diagonal 647, 08028 Barcelona, Catalonia,
Spain \\
%and \\
%{\rm August Romeo}, \\
%{\it Dept de Matem\`atica Aplicada i An\`alisi, \\
%Facultat de Matem\`atiques, Universitat de Barcelona, \\
%Gran Via de les Corts Catalanes 585, 08071 Barcelona, \\ Catalonia. \\

\ms

\renewcommand\baselinestretch{1.4}

\vspace{5mm}

{\bf Abstract}

\end{center}

%Supposing the metric signature of spacetime subject to quantum
%fluctuations, and the compactification of a space dimension,
%we discuss the dynamical determination of signature and
%dimension arising from a
%one-loop effective-potential analysis.
%This leads us to consider the resulting signature of spacetime as
%an effect which is both dynamical and topological.
The formalism of Greensite for treating the spacetime signature as
a dynamical degree of freedom induced by quantum fields is considered,
for spacetimes
with nontrivial topology of the kind ${\bf R}^{D-1} \times {\bf T}^1$,
for varying $D$. It is
shown that a dynamical origin for the Lorentzian signature is possible
in the five-dimensional space ${\bf R}^4 \times {\bf T}^1$ with small
torus radius (periodic boundary conditions), as well as in
four-dimensional
space with trivial topology. Hence, the possibility exists that the
early universe might have been of the Kaluza-Klein
type, \ie multidimensional and of Lorentzian signature.

\newpage

It is well-known that the field equations of general relativity do not
fix the spacetime signature. However, there exist attempts to understand
the signature dynamics or, more specifically, possible signature
transitions from an
Euclidean signature spacetime to a Lorentzian signature spacetime, and
vice-versa, both at a classical and quantum level
\cite{Sakh}-\cite{Der}.
Notwithstanding these efforts, the explanation of the origin of the
Lorentzian signature
of our physical spacetime is still missing.

 Recently, a very interesting
attempt to devise a quantum mechanism for the dynamical origin of the
Lorentzian signature has been made \cite{Gr1,Gr2}.
Let us briefly describe this formalism. Considering
the flat-space metric in the following form:
\beq
\eta_{ab}=\rm{diag}( e^{i\theta}, 1, 1, \dots, 1),
\label{etaab}\eeq
where $\theta\in[ -\pi, \pi]$, one can easily see that the Euclidean
path-integral theory is obtained for the Wick angle $\theta=0$, while
the Lorentzian signature corresponds to $\theta=\pi$.
Starting from this simple setting, it was suggested in ref. \cite{Gr1}
that the Wick angle $\theta$ in
\req{etaab} could be treated as a dynamical degree of freedom, which can
fluctuate in the
interval $\theta\in [-\pi, \pi]$. Then, in order to fix the value of
this degree
of freedom and to show that the Lorentzian signature is in fact the
preferred
choice, the effective potential $V( \theta )$ for $\theta$ has been
calculated in
\cite{Gr1,Gr2}  at one-loop level, under the following assumptions:
(i) For free fields of equal mass, the contributions to the whole
path integral from any propagating bosonic degree of freedom is equal
and inverse to the contribution of the corresponding fermionic
degree of freedom.
(ii) For scalars, the real-valued invariant volume (De Witt measure)
 of integration is used.

Under these conditions, the one-loop potential $V( \theta )$ induced by
a massless scalar in flat spacetime
(\ie $g_{\mu \nu}=e^a_{\mu} \eta_{ab} e^b_{\mu}=\eta_{\mu \nu}$) is
given by \cite{Gr1,Gr2}
\beq V(\theta )=
-{ \log \det^{-1/2} ( - \sqrt{\eta} \eta^{ab} \partial_a \partial_b )
\over
\int d^Dx }, \label{Vlogdet}\eeq
and use of heat-kernel regularization yields
\beq
V(\theta)=-{1 \over 2}\int_{\Lambda}^{\infty} {ds \over s}
\int {d^D p \over (2 \pi )^D}
\exp\left\{\ds -s
[ \alpha p_0^2 + \beta ( p_1^2+\dots +p_{D-1}^2 )] \right\} ,
%\hspace{1cm} \rep \alpha, \rep \beta >0,
\label{hkV}\eeq
where $\Lambda$ is a cutoff, $\alpha=e^{-i {\theta \over 2}}$,
$\beta=e^{i {\theta \over 2}}$. In \cite{Gr2} it is explained why
heat-kernel regularization is the best one to use in this context.
Taking into account assumption (i), the multiplier $(n_B-n_F)$
---where $n_F$($n_B$) is the fermionic (bosonic) number--- appears in
front of \req{hkV}. Finally, using the above formalism it was shown in
Refs.
\cite{Gr1,Gr2} that the Lorentzian signature is uniquely connected with
$D=4$ dimensions, as it is given by the stationary point of the
potential.

Now, our point is the following. Let us start discussing some
modifications of the
above formalism, for it is quite reasonable that the picture above
described
may be valid in the early universe, perhaps between the Quantum Gravity
and the GUTs epochs. However, at this stage of the evolution of the
early universe, its curvature and temperature were still
very strong, and an external electromagnetic field might have existed.
Moreover, most probably
the topology of the universe at this epoch was highly nontrivial.
Surely, all
these effects (and combinations thereof) may change the picture
described in
\cite{Gr1,Gr2}, even qualitatively. In this letter we will restrict
ourselves to consider
only the influence of nontrivial topology on the above formalism.

So, we will suppose that some massless fields, which have induced
the $V(\theta)$ of eq. \req{hkV}, live in a flat spacetime of topology
${\bf R}^{D-1}\times {\bf T}^1$ (for a general discussion of QFT
on topologically nontrivial backgrounds, see \cite{EORBZ}). Then,
expressions \req{Vlogdet} and \req{hkV} still conserve the same
form. However,
now one of the space coordinates, $x_1$, for example, is compactified,
so that the
 corresponding momentum component ---$p_1$--- is discretized
and one has to do replacements of the type
\beq \int {d p_1 \over 2 \pi} \to
{1 \over 2 \pi R} \sum_{n=-\infty}^{\infty}. \eeq
The specific discrete values of $p_1$ will depend on whether the
fields are subject to periodic or antiperiodic boundary conditions.

\ni{\bf (a) Periodic boundary conditions}:
$p_1^2={n^2 \over R^2}$, $n \in {\bf Z}$.

After integrating the non-compactified components of $p$, we find
\beq
V(\theta )=-{\pi^{D-1 \over 2} \over
2R (2 \pi)^D \alpha^{1 \over 2} \beta^{{D\over 2}-1} }
\int_{\Lambda}^{\infty} {ds \over s} \, s^{-{D-1 \over 2}}
\theta_3\left( 0 \left\vert {s \beta \over \pi R^2} \right. \right),
\label{Vper}
\eeq
where
\beq \theta_3(0 | z)\equiv\sum_{n=-\infty}^{\infty} e^{-\pi z n^2}
=1+2\sum_{n=1}^{\infty} e^{-\pi z n^2}.
\label{deftheta3}\eeq

Now, let us consider the different limits of expr. \req{Vper}.

\ni{\bf (a1)} {$R \to 0$}, more specifically,
$\left\vert {\beta \Lambda \over R^2} \right\vert \gg 1$.
That limit corresponds to very strong nontrivial topology.
Thus, we take the expression \req{deftheta3} itself
as a large-$z$ expansion. Integration proceeds term by term
with the help of
\beq \int_z^{\infty} dt \ t^{a-1} e^{-t}= \Gamma( a, z ) , \
\label{igamma} \eeq
where $\Gamma( a, z )$ is an incomplete gamma function \cite{Gra},
whose asymptotic behaviour for large arguments is given by
\beq
\Gamma( a, z ) \sim
z^{a-1} e^{-z}
\left[ 1 + O\left( 1 \over z \right) \right], |z| \gg 1.
\label{Gasymp} \eeq
As a result, the expansion
\beq
V(\theta)=
-{ 2 \over
(4 \pi)^{D+1 \over 2} R \alpha^{1 \over 2} \beta^{{D \over 2}-1}
\Lambda^{D-1 \over 2} }
\left[
{1 \over D-1}
+{1 \over {\beta\Lambda \over R^2}} e^{-{\beta\Lambda \over R^2}}
+O\left(
{1 \over \left( \beta\Lambda \over R^2 \right)^2 }
e^{-{\beta\Lambda \over R^2}}, \dots,
{1 \over { \beta\Lambda \over R^2}} e^{-{4 \beta\Lambda \over R^2}},
\dots
\right)
\right]
\label{persmallR}
\eeq
follows. Note that we have assumed $\rep\alpha >0$, $\rep\beta >0$.

\ni{\bf (a2)}  $R \to \infty$, more precisely
$\left\vert{ \pi^2 R^2 \over \beta \Lambda }\right\vert \gg 1$.
Such a limit corresponds to ``switching off" the compactification.
First, a
small-$z$ expansion of $\theta_3$ must be performed. It can be obtained
by the reciprocal transformation
\beq \theta_3(0|z)=
{1 \over \sqrt{z}}
\theta_3\left( 0 \left\vert {1 \over z} \right. \right) .
\label{recitheta3}\eeq
Then, we can proceed similarly to the previous case, with the difference
that now we are led to make variable changes of the type
$\ds u={\pi^2 R^2 n^2 \over \beta s}$ and split the integration domains
in the way
$\ds \int_0^{{\pi^2 R^2 n^2 \over \beta \Lambda}} du \ =
\int_0^{\infty}du \
- \int_{{\pi^2 R^2 n^2 \over \beta \Lambda}}^{\infty}du $.
Doing so, incomplete $\Gamma$ functions appear again. After using
the asymptotic expansion \req{Gasymp} for large values of their
arguments, one arrives at
\bea
V(\theta)\ds &=&-{1 \over
(4 \pi)^{D/2} \alpha^{1 \over 2} \beta^{D-1 \over 2} \Lambda^{D \over2}}
 \ds \left[ {1 \over D}
+{1 \over \left( \pi^2 R^2 \over \beta \Lambda \right)^{D \over 2}}
\Gamma\left( D \over 2 \right) \zeta(D) \right. \nn
&&\ds\left. -{1 \over { \pi^2 R^2 \over \beta \Lambda }}
e^{-{\pi^2 R^2 \over \beta \Lambda}}
+O\left(
{1 \over \left( \pi^2 R^2 \over \beta \Lambda \right)^2 }
e^{-{\pi^2 R^2 \over \beta \Lambda}}, \dots,
{1 \over { 4 \pi^2 R^2 \over \beta \Lambda } }
e^{-{4 \pi^2 R^2 \over \beta \Lambda}}, \dots
\right) \right] . \nn
\label{perlargeR}\eea
$\zeta$ being the Riemann zeta function. The term where it occurs
yields precisely the result which one would obtain after removing the
$n=0$ piece and performing zeta-function regularization of the rest.
As one can notice, its contribution is ---of course--- independent
of the cutoff $\Lambda$.

\ni{\bf (b) Antiperiodic boundary conditions}:
$p_1^2={\left( n+ {1 \over 2} \right)^2 \over R^2}$, $n \in {\bf Z}$.

Now, the integration of the non-compactified $p$-components yields
\beq
V(\theta)=-{\pi^{D-1 \over 2} \over
2R (2 \pi)^D \alpha^{1 \over 2} \beta^{{D\over 2}-1} }
\int_{\Lambda}^{\infty} {ds \over s} s^{-{D-1 \over 2}}
\theta_2\left( 0 \left\vert {s \beta \over \pi R^2} \right. \right),
\label{Vantiper}
\eeq
where
\beq \theta_2(0|z)\equiv
\sum_{n=-\infty}^{\infty} e^{-\pi z (n+{1 \over 2})^2}
=2\sum_{n=0}^{\infty} e^{-\pi z (n+{1 \over 2})^2} .
\label{deftheta2}\eeq
This last equality can be viewed as a large-$z$ expansion, and will
therefore
be used for calculating the small-$R$ expression in a way analogous to
the corresponding periodic case.

\ni{\bf (b1)}  For $R \to 0$, that is
for
$\left\vert{\beta \Lambda \over R^2}\right\vert \gg 1$, we obtain
\beq
V(\theta)=
-{ 2 \over
(4 \pi)^{D+1 \over 2} R \alpha^{1 \over 2} \beta^{{D \over 2}-1}
\Lambda^{D-1 \over 2} }
\left[
{1 \over {\beta\Lambda \over 4 R^2}} e^{-{\beta\Lambda \over 4 R^2}}
+O\left(
{1 \over \left( \beta\Lambda \over 4 R^2 \right)^2}
e^{-{\beta\Lambda \over 4 R^2}}, \dots,
{1 \over 9{ \beta\Lambda \over 4 R^2 }}
e^{-9 { \beta\Lambda \over 4 R^2}}, \dots \right)
\right].
\label{antipersmallR}
\eeq

\ni{\bf (b2)} For $R \to \infty$,
in order to obtain a small-$z$
expansion from \req{deftheta2} ---which is just the opposite--- this
time we take advantage of the transformation
\[ \theta_2\left( 0 \left\vert z \right. \right)=
{1 \over \sqrt{z} }
\theta_4\left( 0 \left\vert {1 \over z} \right. \right), \]
\beq \theta_4(0|z)\equiv
\sum_{n=-\infty}^{\infty}(-1)^n e^{-\pi z n^2}
=1+2\sum_{n=1}^{\infty}(-1)^n e^{-\pi z n^2} . \eeq
This is a useful relation, as well as its counterpart \req{recitheta3}
for the analogous periodic case, and both can
be regarded as  special forms of the general reciprocal transformation
for $\theta$ functions (see \eg \cite{epsma}).
%It is now easy
%to use the expression of $\theta_4$ as an expansion for small argument.
After integrating, one gets
\bea
V(\theta)&=&-{1 \over
(4 \pi)^{D/2} \alpha^{1\over 2} \beta^{D-1 \over 2} \Lambda^{D \over 2}}
 \ds \left[ {1 \over D}
-{1 \over \left( \pi^2 R^2 \over \beta \Lambda \right)^{D/2}}
\Gamma\left( D \over 2 \right) \eta(D) \right. \nn
&&\ds \left. +{1 \over { \pi^2 R^2 \over \beta \Lambda }}
e^{-{\pi^2 R^2 \over \beta \Lambda}}
+O\left(
{1 \over \left( \pi^2 R^2 \over \beta \Lambda \right)^2 }
e^{-{\pi^2 R^2 \over \beta \Lambda}}, \dots,
{1 \over { 4 \pi^2 R^2 \over \beta \Lambda }}
e^{-{4 \pi^2 R^2 \over \beta \Lambda}}, \dots
\right)
\right] , \nn
\label{antiperlargeR}
\eea
which is valid for $\left\vert{\pi^2 R^2 \over \beta \Lambda}\right\vert
\gg 1$. Here $\eta$ is the well-known Dirichlet series
\beq \eta(z)\equiv \sum_{n=0}^{\infty} (-1)^{n-1} n^{-z}
=(1-2^{1-z})\zeta(z). \eeq

Now we discuss the physical consequences of the results just
obtained. As we could see, the potential $V(\theta )$ is complex, so we
will
look for the value of $\theta$ determined by the following conditions
\cite{Gr1}:
\beq
\begin{array}{lll}
(i) & \imp V&\mbox{must be stationary, and} \\
(ii) & \rep V&\mbox{must be a minimum.}
\end{array}
\label{condi}
\eeq

Note also that, in accordance with the first assumption of \cite{Gr1},
our results \req{Vper}, \req{persmallR}, \req{perlargeR},
\req{Vantiper}, \req{antipersmallR} and \req{antiperlargeR}
should be multiplied by $(n_B-n_F)$.

We are ready to start the analysis of the effective potential.
For the
$R \to \infty$ limit (trivial topology), the analysis made in \cite{Gr1}
shows that there exists a unique solution
\beq n_F>n_B, \ \ \theta=\pm\pi, \ \ D=4. \label{solGr}\eeq
(Note that for $D=2$ or $n_F=n_B$ there is no preferable choice of
$\theta$, as $V$ ceases to depend on it. We shall call these cases
`special'). The solution \req{solGr}
leads the authors of \cite{Gr1,Gr2} to conclude that the Lorentzian
signature is chosen by quantum dynamics {\it only} in $D=4$.

Now we may consider the case when the radius of the compactified
dimension is small, and quantum fields satisfy periodic boundary
conditions. Then, using the leading term of \req{persmallR}
(with the multiplier $(n_B-n_F)$)
we  find the following unique solution satisfying the requirements
\req{condi} (apart from the `special' cases, which are here $D=3$ or
$n_F=n_B$)
\beq n_F>n_B, \ \ \theta=\pm\pi, \ \ D=5. \eeq
Thus, the Lorentzian signature is singled out {\it also} in $D=5$, but
only if
the fifth dimension is compactified with a very small radius. Therefore,
the formalism of the dynamical degree of freedom associated to the Wick
angle provides a window for Kaluza-Klein-type theories.
It is not difficult to show that, had we started from the topology
${\bf R}^{D-n} \times {\bf T}^n \ ( n<D-1)$, and did a similar
small-$R$ expansion (taking all the torus radii to be equal) we would
have found the following result:
\beq n_F>n_B, \ \ \theta=\pm\pi, \ \ D=4+n. \eeq
Hence, an early universe with a nontrivial topology could have been
multidimensional, and this is compatible with the Lorentzian signature.

As a last example, let us consider the potential \req{antipersmallR}
corresponding to small radius for the compactified dimension and
antiperiodic boundary conditions for the quantum fields.
Then, from \req{antipersmallR} we have
\beq
V(\theta) \sim
{ 8(n_F-n_B) R \ e^{-{\beta \Lambda \over 4 R^2}} \over
(4 \pi)^{D+1 \over 2} \alpha^{1 \over 2} \beta^{D \over 2}
\Lambda^{D+1 \over 2} } .
\label{LTantipersmallR}
\eeq
The conditions \req{condi} for the $V(\theta)$ in \req{LTantipersmallR}
mean that there should be some value of $\theta$, say $\bar\theta$,
simultaneously satisfying
\beq
\begin{array}{c}
\begin{array}{ll}
\ds (n_B-n_F) e^{ -{\Lambda \over 4R^2} \cos{\theta \over 2} }
&\ds\left[ {\Lambda \over 8R^2} \sin{\theta \over 2} \
\sin\left(
{D-1 \over 4}\theta + {\Lambda \over 4R^2} \sin{\theta \over 2}
\right) \right. \\
&\ds\left.\left. +\left( {D-1 \over 4}
+{\Lambda \over 8R^2}\cos{\theta \over 2} \right)
\cos\left(
{D-1 \over 4}\theta + {\Lambda \over 4R^2} \sin{\theta \over 2}
\right)
\right] \right\vert_{\theta=\bar\theta}=0,
\end{array} \\
%\ds {\rm min}_{\theta | -\pi \le \theta \le \pi}
\ds  (n_F-n_B)e^{-{\Lambda \over 4R^2}\cos{\theta \over 2} }
\cos\left(
{D-1 \over 4}\theta + {\Lambda \over 4R^2} \sin{\theta \over 2}
\right) \ \ \
\mbox{has a minimum at $\theta=\bar\theta$}.
\end{array}
\label{condii}
\eeq
In other words, one takes the potential \req{LTantipersmallR} and
requires the
coincidence of stationary points of its imaginary part with minima of
its real part, bearing in mind all the time that we are restricted to
$\theta\in [-\pi,\pi]$.
Now, this situation differs from the previous cases in the variable
nature of this $V(\theta)$ as ${\Lambda \over 4R^2}$ varies.
We have plotted the curves representing its rescaled real and imaginary
parts
for $D=4,5$ and for different values of $b\equiv {\Lambda \over 4R^2}$
(see Fig. 1).
The behaviour observed is drastically modified as this parameter
increases, going from a regime of noticeable oscillation to
one in which a plateau around the origin is formed by both $\rep V$
and $\imp V$.
After studying $V$ for other values of $b$ not shown in the figure,
we have detected that the width of this plateau increases as $b$ grows.
The flatness of this region
indicates an angular range where $V$ becomes practically independent
of $\theta$, and therefore,
within this range no particular spacetime signature is preferred.
(Notice that the situation where $b \gg 1$ is precisely
the one in which \req{LTantipersmallR} is an acceptable approximation
for $V(\theta)$).
Apart from this, and generally speaking, there is no genuine
coincidence of stationary points and minima, unless very specific
values of $b$ and $D$ are deliberately chosen for that purpose.

Summing up, we have discussed the possibility of a dynamical origin of
the
Lorentzian signature in a universe with nontrivial topology. Using very
precise mathematical techniques, we
have shown that the Lorentzian signature can actually be dynamically
induced
in a multidimensional universe with the topology ${\bf R}^4 \times {\bf
T}^n$, where the radii of the torus are small. It would certainly be of
interest to estimate the mass effects \cite{Gr2} on the above results,
since it was already shown in \cite{Gr2} that considering
massive
fields might increase the resulting $D$ until $D=6$. Hence, the
combination of both effects (topology and nonzero mass) may lead
to changes in the
above results. Finally, let us note that it would also be of great
interest
to understand the influence of quantum gravity on the dynamical origin
of the spacetime signature. This subject surely deserves further study.

\vskip3ex

\ni{\Large \bf Acknowledgements}

We would like to thank Robin Tucker for interesting correspondence. This
work has been supported by
DGICYT (Spain), project no. PB90-0022, and by
CIRIT (Generalitat de Catalunya).

\newpage

\ni{\Large \bf Figure captions}

\ni{\bf Figure 1}.
Curves representing $\rep v(\theta)$ ---solid line---
and $\imp v(\theta)$ ---dashed line--- with $v(\theta)$
denoting the {\it rescaled} potential
$e^{-b \cos{\theta \over 2}
+i\left( a \theta+ b\sin{\theta \over 2} \right) }$,
with $a\equiv{D-1 \over 4}$, $b\equiv{\Lambda \over 4 R^2}$
for $-\pi \le \theta \le \pi$.
The plots correspond to
$D=4, D=5$ and to two different values of $b$:
(a) $D=4, b=1$, (b) $D=4, b=10$, (c) $D=5, b=1$, (d) $D=5, b=10$.
In (b) and (d) the formation of a wide plateau around the origin for
large values of $b$ is already apparent.

\end{document}